\documentclass[final,1p,times]{elsarticle}

\usepackage{amssymb}
\usepackage[section] {placeins}
\usepackage[all]{xypic}
\usepackage{epsfig}
\usepackage{graphicx}
\usepackage{color}
\usepackage{dcolumn}
\usepackage{bm}
\usepackage{mdframed}
\usepackage{marginnote}
\usepackage{float}
\usepackage{lineno}
\usepackage{longtable}
\usepackage{rotating}
\usepackage{array}
\usepackage{multicol}
\usepackage{multirow}

\usepackage{xcolor}
\usepackage{tikz}
\usepackage{multirow}

\newcommand{\bp}{{\mathbf p}}
\newcommand{\bq}{{\mathbf q}}
\setcitestyle{square}

\usepackage{booktabs}

\journal{Physics Letters B}

\begin{document}

\begin{frontmatter}



\title{Heavy Tetraquarks in the Diquark-Antidiquark Picture}


\author[label1]{M. R. Hadizadeh}
\ead{hadizadm@ohio.edu}

\author[label2]{A. Khaledi-Nasab}
 \ead{ak705714@ohio.edu}
 
 \address[label1]{Institute of Nuclear and Particle Physics and Department of Physics and Astronomy, Ohio University, Athens, OH 45701, USA,}
 \address[label2]{Department of Physics and Astronomy, Ohio University,
Athens, OH 45701, USA}

\begin{abstract}
The homogeneous Lippmann-Schwinger integral equation is solved in momentum space to calculate the masses of heavy tetraquarks with hidden charm and bottom. The tetraquark bound states are studied in the diquark-antidiquark picture as a two-body problem. A regularized form of the diquark-antidiquark potential is used to overcome the singularity of the confining potential at large distances or small momenta. Our numerical results indicate that the relativistic effect leads to a small reduction in the mass of heavy tetraquarks, which is less than $2\,\%$ for charm and less than $0.2\,\%$ for bottom tetraquarks.
The calculated masses of heavy tetraquarks for $1s$, $1p$, $2s$, $1d$ and $2p$ states are in good agreement with other theoretical calculations and experimental data. Our numerical analysis predict the masses of heavy tetraquarks for $3s$, $2d$ and $3p$ states for the first time, and we are not aware of any other theoretical results or experimental data for these states.

\end{abstract}


\end{frontmatter}

\section{Introduction}
In the early 60's, quarks become the constituent of strong interaction and it raise to be a handy tool to describe the observed particles in the hadron spectrum \cite{gell1964schematic}. It is known that quarks can be in groups like a system of coupled quarks and coupled antiquarks. The idea of strongly coupled two-quarks-two-antiquarks mesons to baryon-antibaryon channels was suggested by R. Jaffe \cite{PRD}, where the MIT bag model used to predict the quantum numbers and the masses of prominent states. Afterword a non-relativistic potential model (NRPM) was presented by Zouzou et al. \cite{zouzou1986four} to investigate the system consisting of two-quarks and two-antiquarks with equal/unequal masses. Precisely they were searching for probable bound states under the threshold for the spontaneous dissociation into two-mesons system. Relativistic quark-antiquark bound-state by considering the spin-dependent interactions in momentum space  has been explored by Jean et al. \cite{jean1994relativistic} where it was the first study toward a relativistic three-quark bound-state using a Hamiltonian consistent with the Wigner-Bargmann theorem and macroscopic locality.
They used a potential which includes confinement and is of the general form consistent with rotation, space-reflection, and time-reversal invariance and it was a combination of linear, Coulomb, spin-spin, spin-orbit, and tensor terms. An investigation on heavy-light tetraquark bound states by means of a chiral constituent quark model reported by Vijande et al. \cite{vijande2004tetraquarks}. 
They also presented the hyperspherical harmonic formalism for tetraquarks and they studied the systems made of quarks and antiquarks of the same flavor \cite{vijande2007hyperspherical}. 

In a series of studies by Ebert et al. \cite{ebert2006masses, ebert2009relativistic,ebert2011masses}, they have proposed a relativistic model of the ground and excited states of heavy tetraquarks for hidden charm and bottom within the diquark-antidiquark ($D\bar{D}$) picture (heavy-light diquark and antidiquark). They have treated the light quarks, in the heavy-light diquark, and diquarks quite relativistically. 
Additionally they have discussed the  experimental data on charmonium-like states above open charm threshold and they have found that the  masses of ground state tetraquarks with hidden bottom are below the open bottom threshold. 
They have also shown that the anomalous scalar $D^{*}_{s0} (2370)$ and axial vector $D_{s1}(2460)$ mesons cannot be considered as $D\bar{D}$ bound states, while $D_s(2632)$ and $D^{*}_{sJ}(2860)$ could be interpreted as scalar and tensor tetraquarks, respectively.

In a recent paper by Monemzadeh et al. \cite{Monemzadeh-PLB741}, the tetraquark masses are calculated in configuration space using two-body bound state of diquark-antidiquark. They have solved the spin-independent non-relativistic LS equation only for heavy charm tetraquarks restricted to $s-$wave channel.  
However we have questioned the validity of their results and we brought the criticism in a comment on this paper \cite{Hadizadeh-EPJC75}.
The main goal of the comment was to show it remains completely unclear, how the authors of Ref. \cite{Monemzadeh-PLB741} can discriminate between the masses of tetraquarks with axial-vector diquark content and different total angular momentum $J$ in a spin-independent framework. Also it has been shown that the paper suffers from few computational issues, for instance their regularization cutoff is not high enough to achieve  accurate results.

In this letter we have solved the non-relativistic and relativistic homogeneous Lippmann-Schwinger integral equation using a regularized form of the spin-independent $D\bar{D}$ potential in momentum space. The tetraquark bound states are studied as a two-body problem in the $D\bar{D}$ picture and the masses of heavy tetraquarks with hidden charm and bottom are calculated. The role of the relativistic effect in the mass spectrum of tetraquarks is studied in detail.

\section{Tetraquark bound states in the diquark-antidiquark picture in momentum space}

The relativistic bound state of $D\bar{D}$ system with the relative momentum of $\bp$ in a partial wave representation is given by
\begin{equation}
\label{eq.LS}
\psi_{l} (p) = \frac{1}{m_T- \omega(p)} \, \int_{0} ^{\infty} dp' \, p'^2 \, V_l (p,p') \,
\psi_{l} (p'),
\end{equation}
where $\omega(p)=\sqrt{m_D^2+p^2}+\sqrt{m_{\bar D}^2+p^2}$. $m_T$, $m_D$ and $m_{\bar D}$ are the masses of tetraquark, diquark and antidiquark, correspondingly. $V_l (p,p')$ is the projection of the potential $V(\bp,\bp') \equiv V(p,p',x)$ in the partial wave channel $l$
\begin{equation}
\label{eq.Vl}
V_l (p,p') = 2 \pi \int_{-1} ^{+1} dx \, P_l(x) \, V(p,p',x).
\end{equation}
In the non-relativistic limit the free propagator $[m_T- \omega(p)]^{-1}$ is replaced by $(E-\frac{p^2}{2\,\mu_{D\bar{D}}})^{-1}$, where $E=m_T-m_D - m_{\bar{D}}$ is $D\bar{D}$ binding energy and $\mu_{D\bar{D}}= \frac{m_{D} m_{\bar{D}}}{m_D +m_{\bar{D}}}$ is the reduced mass of $D\bar{D}$ system.
In this study the spin-independent part of heavy $D\bar{D}$ potential of Ref. \cite{PhysRevD.76.114015} is used
\begin{equation}
 \label{eq.V_dd}
V (r) = V_{Coul} (r) + V_{conf} (r),
\end{equation}
with the linear confining
\begin{equation}
 \label{eq.V2}
V_{conf} (r) = Ar+B,
\end{equation}
and the Coulomb-like one-gluon exchange potential  
\begin{equation}
 \label{eq.V2}
V_{Coul} (r) = \gamma \, \frac{F_{D} (r) F_{\bar{D}} (r)}{r},  \quad  \gamma = \frac{-4}{3} \alpha_s.
\end{equation}
$F_D$ and $F_{\bar{D}}$ are the form factors of diquark and antidiquark, correspondingly, and have the following functional form
\begin{equation}
 \label{eq.F}
F (r)= 1-e^{\alpha r -\beta r^2}.
\end{equation}
By considering this functional structure of the form factors, the Coulomb part of the $D\bar{D}$ potential can be rearranged as 
\begin{equation}
\label{eq.Vcoul}
V_{Coul} (r)=\frac{\gamma}{r}  \left (1+ \sum\limits_{j=1}^3 (-)^{j} e^{-\alpha_j r -\beta_j r^2} \right ),
\end{equation}
where $\alpha_j$ and $\beta_j$ are defined by diquark and antidiquark form factor parameters, as shown in Table \ref{Table_alpha_beta}.
The parameters of this model are fixed from the analysis of heavy quarkonia masses and radiative decays \cite{Galkin-SJNP44,Galkin-SJNP51,Galkin-SJNP55}.
The confining parameters are $A=0.18$ GeV$^2$ and $B=-0.30$ GeV which have standard values of quark models. The strong coupling constant $\alpha_s$ is given by \cite{ebert2011masses}
\begin{equation}
\label{alphas}
\alpha_s (\mu)= \frac{4\pi}{\beta_{\alpha_s}} \frac{1}{\ln \left ({\frac{4\mu^2_{D\bar{D}} + M_{\alpha_s}^2}{\Lambda^2}} \right)} , \quad \beta_{\alpha_s}=11- \frac{2}{3} n_f, \quad
M_{\alpha_s}= 2.24 \sqrt{A}, \quad \Lambda=0.413~ \text{GeV},
\end{equation}
where $n_f=4$ is the number of flavor quarks.
In our calculations we have used the masses of diquark (antidiquark) and form factor parameters of Ref. \cite{ebert2006masses}, which are given in Table \ref{Table.pot_parameters}.

\begin{table}[hbt] 
\centering
\begin{tabular}{cccccc}
\toprule
j & $\alpha_j$ & $\beta_j$ \\
\hline
1 & $\alpha_D$ & $\beta_D$ \\
2 & $\alpha_D + \alpha_{\bar{D}}$ & $\beta_D + \beta_{\bar{D}}$ \\
3 & $\alpha_{\bar{D}}$ & $\beta_{\bar{D}}$ \\
\bottomrule
\end{tabular}
\caption{The form factor parameters $\alpha_j$ and $\beta_j$ in Coulomb-like potential of Eq. (\ref{eq.Vcoul}).}
\label{Table_alpha_beta}
\end{table}

\begin{table}[hbt] 
\centering
\begin{tabular}{cccccccccccc}
\toprule
Tetraquark content & $D\bar{D}$ type &  $m_D=m_{\bar{D}}$ (MeV)&$\alpha_D=\alpha_{\bar{D}}$ (GeV)&$\beta_D=\beta_{\bar{D}}$ (GeV$^2$)      \\ \hline
\multirow{2}{*}{$cq\bar{c}\bar{q}$} & $S\bar{S}$ & 1973& 2.55 &0.63 \\ \cmidrule{2-5}  
& $A\bar{A}$ & 2036& 2.51 &0.45 \\ 
\hline
\multirow{2}{*}{$cs\bar{c}\bar{s}$}& $S\bar{S}$ & 2091& 2.15 & 1.05 \\  \cmidrule{2-5} 
 & $A\bar{A}$ & 2158&2.12& 0.99 \\ 
\hline
\multirow{2}{*}{$bq\bar{b}\bar{q}$}& $S\bar{S}$ &  5359 &6.10 & 0.55 \\ \cmidrule{2-5} 
& $A\bar{A}$ & 5381& 6.05 &0.35 \\
\hline
\multirow{2}{*}{$bs\bar{b}\bar{s}$}& $S\bar{S}$ & 5462 & 5.70 &0.35 \\ \cmidrule{2-5} 
& $A\bar{A}$ & 5482 & 5.65 &0.27 \\
\bottomrule
\end{tabular}
\caption{The masses of diquark and antidiquark ($m_D$ and $m_{\bar{D}}$) and the form factor parameters ($\alpha_D, \alpha_{\bar{D}}, \beta_D,\beta_{\bar{D}}$) of heavy-light diquarks. $S$ and $A$ denote the scalar and axial vector diquarks.}
\label{Table.pot_parameters}
\end{table}

Since the confining part of the $D\bar{D}$ potential is unbounded at large distances, it leads to a singularity in the integral equation (\ref{eq.LS}) at small momenta.
To overcome this singularity one can use the regularized form of the confining potential \cite{Hadizadeh_AIP1296}. To this aim one can keep the divergent part of the potential fixed after exceeding a certain distance. This procedure creates an artificial barrier and the influence of tunneling barrier is manifested by significant changes in the energy eigenvalues at small distances.
By following this strategy and keeping the potential fixed at $r_c$, the Fourier transformation of the regularized form of the potential in momentum space is given by
\begin{eqnarray}
\label{eq.FTV}
V(p,p',x)  &=&  
V_0 \, \delta^3(\bq)  \cr
&-& \frac{V_0}{2\pi^2 q} \left (-\frac{r_c}{q} \cos(qr_c)+\frac{1}{q^2} \sin(qr_c) \right)\cr
&+&\frac{A}{2\pi^2 q} \left (\frac{2}{q^3} \cos(qr_c) + \frac{2r_c}{q^2} \sin(qr_c) -\frac{r_c^2}{q} \cos(qr_c) -\frac{2}{q^3} \right) \cr
&+& \frac{B}{2\pi^2 q} \left (\frac{1}{q^2} \sin(qr_c)-\frac{r_c}{q} \cos(qr_c) \right) \cr
&+& \frac{\gamma}{2\pi^2 q} \left (\frac{1}{q}- \frac{1}{q} \cos(qr_c) \right)\cr
&+& \frac{\gamma}{2\pi^2 q}\sum\limits_{j=1}^3 (-)^j Im \left [ e^{\beta_j r_j^2} \sqrt{\frac{\theta_f -\theta_i}{-2\beta_j}
\left(e^{-\beta_j R_f^2} - e^{-\beta_j R_i^2} \right)} \right],
\end{eqnarray}
where 
\begin{eqnarray}
\label{eq.FTV}
q &=& |\bq|= |\bp- \bp'|=\sqrt{p^2+p'^2-2pp'x}, \cr
V_0 &=& Ar_c +B+ \frac{\gamma}{r_c} \left (1+\sum\limits_{j=1}^3 (-)^{j} e^{-\alpha_j r_c -\beta_j r_c^2} \right), \cr
R_i &=& \sqrt{2} \, r_j , \cr
R_f &=&\sqrt{2} \, (r_j+r_c)  , \cr
\theta_i &=& tan^{-1} \left (\frac{r_j}{r_j+r_c} \right), \cr
\theta_f &=& tan^{-1} \left(\frac{r_j+r_c}{r_j} \right) , \cr
r_j &=& \frac{iq-\alpha_j}{-2\beta_j}.
\end{eqnarray}

\section{Results and Discussion} \label{results}

\begin{figure}[H] 
  \centering
    \includegraphics[width=.8\textwidth]{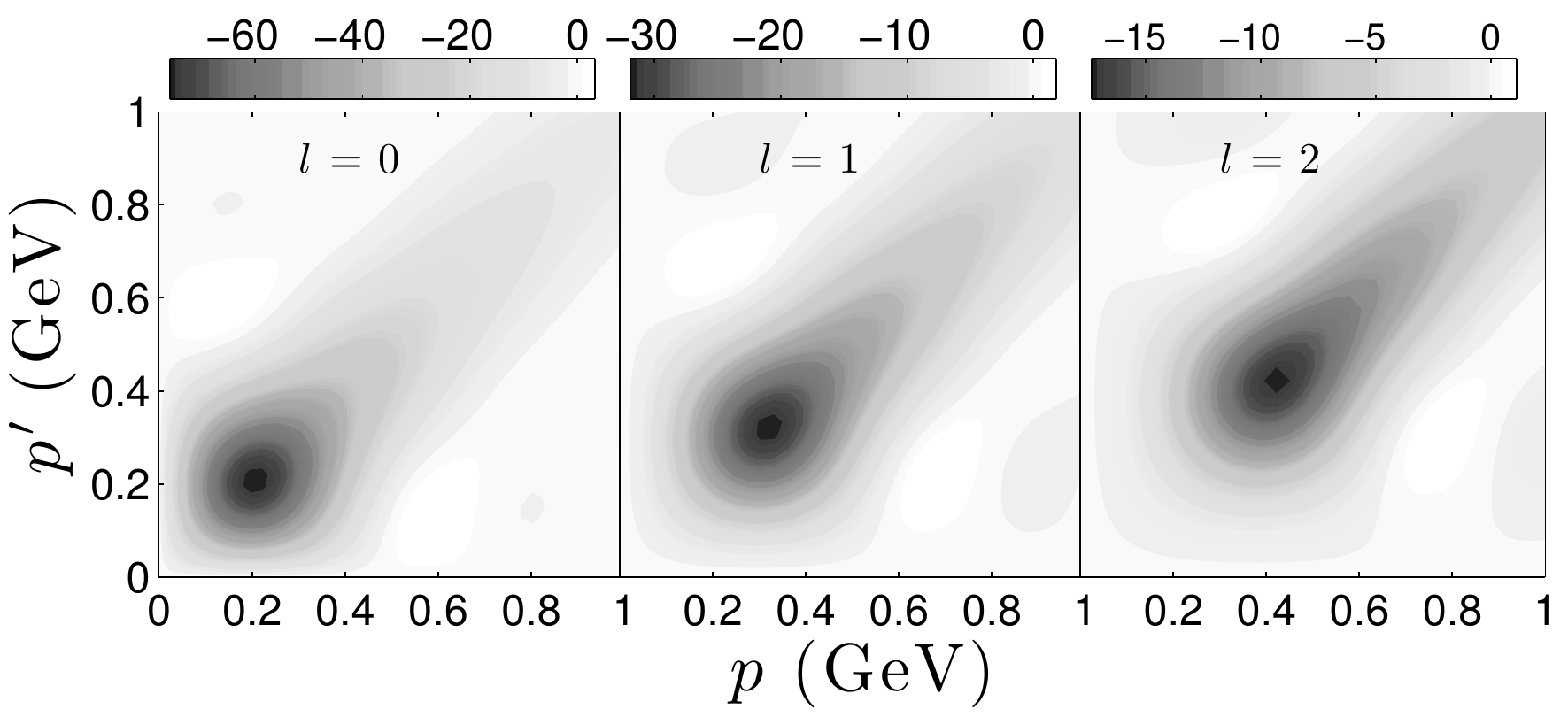} \\
 \caption{The matrix elements of the diquark-antidiquark potential in units of GeV$^{-2}$ for $s-$, $p-$ and $d-$wave channels calculated for $S\bar{S}$ state in the parameterization of $cq\bar{c}\bar{q}$ tetraquark.}    
 \label{Fig.VSS}
\end{figure}

\begin{figure}[H] 
  \centering
    \includegraphics[width=.8\textwidth]{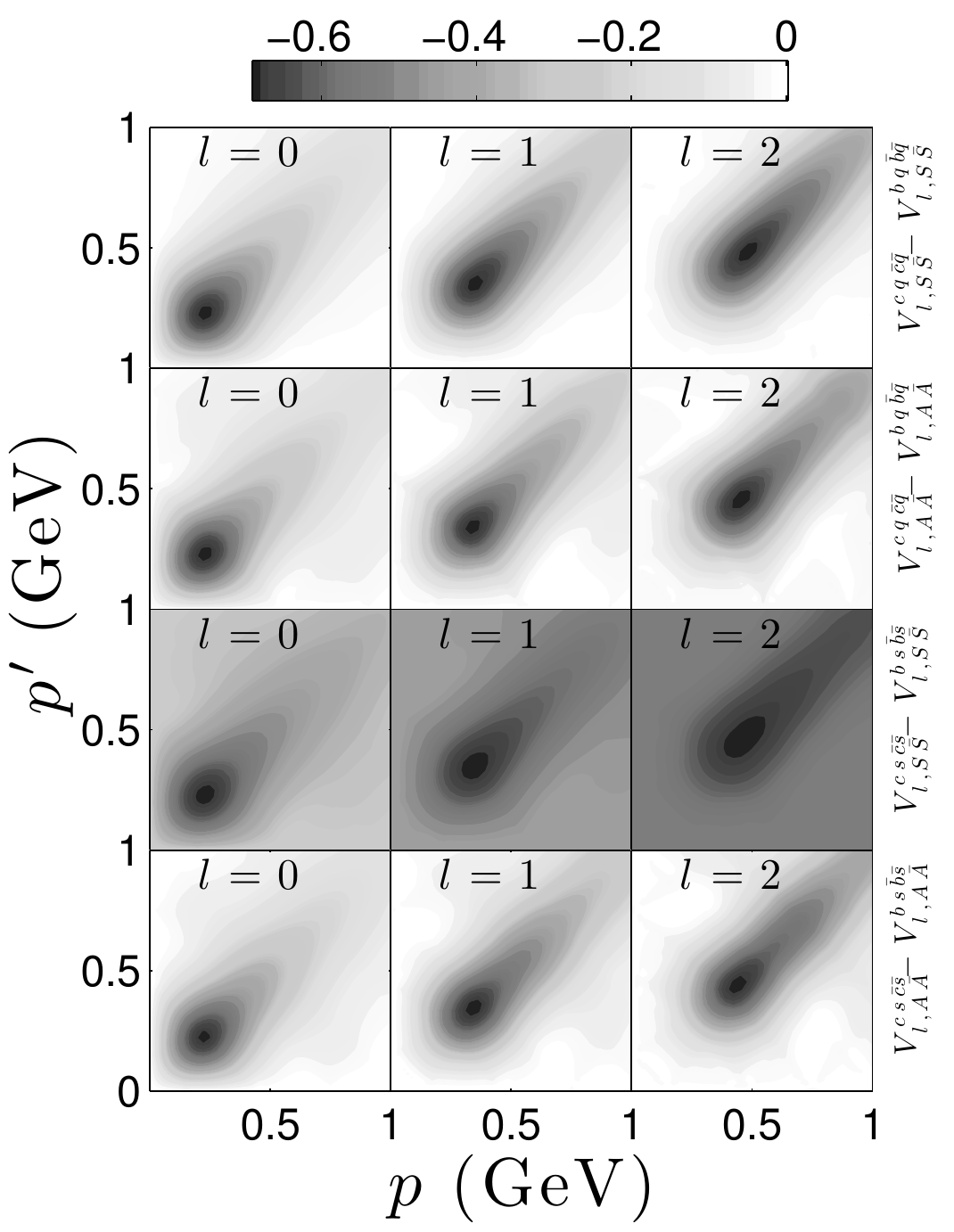} \\
 \caption{The difference of the matrix elements of diquark-antidiquark potential in units of GeV$^{-2}$ for $s-$, $p-$ and $d-$wave channels calculated for $S\bar{S}$ and $A\bar{A}$ states in the parameterization of charm and bottom tetraquarks.}    
 \label{Fig.V_diff}
\end{figure}

In Fig. \ref{Fig.VSS} we have shown an example of the matrix elements of $D\bar{D}$ potential for partial wave channels $s$, $p$ and $d$ in parameterization of $S\bar{S}$ state of $cq\bar{c}\bar{q}$ tetraquark. As it is shown the negative dip of the potential shifts to higher momenta for higher partial waves and its depth becomes smaller with a factor of about 2 for two successive partial waves. The structure of the matrix elements of $D\bar{D}$ potential for $S\bar{S}$ and $A\bar{A}$ states of charm and bottom tetraquarks is similar, but they have a small difference which is shown in Fig. \ref{Fig.V_diff}.

The first step toward the numerical solution of the integral equations (\ref{eq.LS}) and (\ref{eq.Vl}) is discretization of continuous momentum and angle variables and to this aim we have used Gauss-Legendre quadratures.
 The momentum integration interval $[0,\infty)$ is covered by a combination of hyperbolic and linear transformations of Gauss-Legendre points from the interval $[-1,+1]$ to the intervals $[0,p_1] \, + [p_1,p_2] \, + [p_2,p_3]$ as
 \begin{eqnarray}
 \label{eq.mapping}
p_{hyperbolic} = \frac{1+x}{\frac{1}{p_1} + (\frac{2}{p_2}-\frac{1}{p_1} )\, x }, \quad p_{linear}=\frac{p_3-p_2}{2}\, x + \frac{p_3+p_2}{2}.
\end{eqnarray}
The typical values for $p_1, \, p_2$ and $p_3$ in our calculations are $0.5, \, 1$ and $10$ GeV. 
The integral equation (\ref{eq.LS}) can be written schematically as eigenvalue equation $\lambda \, \psi = K(M)\, \psi$, where the physical tetraquark mass $m_T$ is corresponding to eigenvalue $\lambda=1$. The eigenvalue equation can be solved by direct method. For calculation of the masses of charm (bottom) tetraquarks we have solved the integral equation by searching in a wide range of tetraquark mass in the region $3.7 \le M \le 5.4$ GeV ($10.4 \le  M \le 11.7$ GeV) and we have extracted the physical bound states for $\lambda=1$ with a relative error of $10^{-10}$. 

Our numerical results for the masses of charm ($cq\bar{c}\bar{q}$ and $cs\bar{c}\bar{s}$) and bottom ($bq\bar{b}\bar{q}$ and $bs\bar{b}\bar{s}$) tetraquarks for $s-$, $p-$ and $d-$wave channels with total spin ${\cal S}=0$ are listed in Tables \ref{Table.cqcq_cscs} and \ref{Table.bqbq_bsbs}. The tetraquark masses are calculated for scalar $S\bar{S}$ and axial-vector $A\bar{A}$ diquark-antidiquark contents.

\begin{table}[hbt] 
\centering
\begin{tabular}{cccccccccccc}
\toprule
state &  \multicolumn{4}{c}{$cq\bar{c}\bar{q}\,(S\bar{S})$} &   & \multicolumn{4}{c}{$cq\bar{c}\bar{q}\,(A\bar{A})$}  \\  \cmidrule{2-5}  \cmidrule{7-10}
    &   NR   & R  & EFG~\cite{ebert2006masses,Ebert_EPJC58}  &  $\Delta (\%)$   & &   NR   & R  & EFG~\cite{ebert2006masses,Ebert_EPJC58}  &  $\Delta (\%)$ \\ \hline
$1s$   &  3.792  &  3.739 & 3.812  & 1.9 & & 3.919  &  3.869&  3.852 & 0.4  \\
$1p$   &  4.262  &  4.231 & 4.244 & 0.3 &  &4.374 &  4.346  &4.350 & 0.1 \\
$2s$   &  4.419  & 4.357 & 4.375  & 0.4 & &4.535 & 4.469&  4.434 & 0.8 \\
$1d$   &  4.556  &  4.526 & 4.506 & 0.4 &&4.668   &  4.637 & 4.617 & 0.4 \\
$2p$   &  4.697  & 4.644 & 4.666 & 0.5 & &4.816  &  4.771 &  4.765 & 0.1 \\
$3s$   & 4.843   & 4.757  &  & & & 4.944   &  4.862 & \\
$2d$   & 4.933   & 4.876 & & & &  5.037  &  4.983 &  \\
$3p$   & 5.062    &  4.990 & & & &  5.184   & 5.114 & \\
\hline
state &   \multicolumn{4}{c}{$cs\bar{c}\bar{s}\,(S\bar{S})$} &   & \multicolumn{4}{c}{$cs\bar{c}\bar{s}\,(A\bar{A})$}  \\  \cmidrule{2-5}  \cmidrule{7-10}
    &   NR   & R  & EFG~\cite{ebert2006masses,Ebert_EPJC58}   &  $\Delta (\%)$ &     &   NR   & R  & EFG \cite{ebert2006masses,Ebert_EPJC58}  &  $\Delta (\%)$ \\ \hline
 $1s$  &  4.011   &  3.946  &  4.051 & 0.8 &&  4.139    &  4.078   &   4.110 &  0.8 \\
$1p$  &   4.490    & 4.464   &  4.466 & 0.0 &&   4.616    &  4.591  &  4.582 &  0.2  \\
$2s$  &   4.620    &   4.558  &  4.604  & 1.0 &&    4.744   &   4.687  &  4.680 &0.1  \\
$1d$  &   4.770   &   4.743  &  4.728  & 0.3 &&   4.894    &  4.869  &  4.847 & 0.4 \\
$2p$  &  4.920    & 4.870   &  4.884 & 0.3 &&   5.041    &  4.993   &   4.991  & 0.0 \\
$3s$  &  5.039   &   4.964  &  &&&   5.160 &   5.090  &     \\
$2d$  &   5.143  &   5.094  &  &&&   5.263  &  5.216   &    \\
$3p$  &  5.276    &  5.204  &   &&&  5.394    & 5.324   & \\
\bottomrule
\end{tabular}
\caption{Masses of charm diquark-antidiquark states in units of GeV, calculated from non-relativistic (NR) and relativistic (R) Lippmann-Schwinger equation. $\Delta$ is the absolute value of the relative percentage difference between our findings and EFG results.}
\label{Table.cqcq_cscs}
\end{table}

\begin{table}[hbt] 
\centering
\begin{tabular}{cccccccccccc}
\toprule
state &  \multicolumn{4}{c}{$bq\bar{b}\bar{q}\,(S\bar{S})$} &  &  \multicolumn{4}{c}{$bq\bar{b}\bar{q}\,(A\bar{A})$}  \\  \cmidrule{2-5}  \cmidrule{7-10}
    &   NR   & R  &  EFG~\cite{ebert2006masses,Ebert_MPLA24}   &  $\Delta (\%)$ &    &   NR   & R  &  EFG~\cite{ebert2006masses,Ebert_MPLA24} &  $\Delta (\%)$ \\ \hline
$1s$   &  10.426  &  10.410   &   10.471 & 0.6 &&   10.469    &  10.453   &    10.473 & 0.2 \\
$1p$   & 10.813    &  10.806   & 10.807    & 0.0 &&  10.856   &  10.850   &  10.850  & 0.0 \\
$2s$   & 10.914   & 10.899    & 10.917   & 0.2 && 10.958   &   10.942  &  10.942  & 0.0 \\
$1d$   & 11.034  & 11.028 &      11.021  & 0.1 && 11.077  & 11.071 &      11.064    & 0.1 \\
$2p$   & 11.140   &  11.128    &   11.122  & 0.0 &&  11.183  &  11.171   &  11.163  & 0.1 \\
$3s$   &  11.230  &   11.211  &    &&&  11.273   &   11.254 &    \\
$2d$   & 11.310   &  11.299   &    &&&  11.354  &  11.342   & \\
$3p$   &  11.406  &   11.389  &   &&&  11.450  &  11.433   &     \\
\hline
state &   \multicolumn{4}{c}{$bs\bar{b}\bar{s}\,(S\bar{S})$} &  & \multicolumn{4}{c}{$bs\bar{b}\bar{s}\,(A\bar{A})$}  \\  \cmidrule{2-5}  \cmidrule{7-10}
    &   NR   & R  & EFG~\cite{ebert2006masses,Ebert_MPLA24}   &  $\Delta (\%)$ &  &   NR   & R  &  EFG~\cite{ebert2006masses,Ebert_MPLA24}  &  $\Delta (\%)$ \\ \hline
 $1s$   & 10.629  &  10.613  & 10.662   & 0.5 && 10.668    &  10.653  & 10.671  &  0.2\\
$1p$   &  11.015   & 11.009  & 11.002   & 0.1 && 11.054    & 11.048    &   11.039   &  0.1 \\
$2s$   & 11.116  &  11.100   &   11.111  & 0.1 && 11.155   &   11.139  & 11.133     & 0.0 \\
$1d$   & 11.235  & 11.229&     11.216   & 0.1 && 11.274 &   11.268  &   11.255   &  0.1 \\
$2p$   & 11.340   & 11.329    &  11.316   & 0.1 && 11.379   &  11.368   &  11.353   & 0.1 \\
$3s$   &  11.430  &   11.411  &   & && 11.469   &  11.428   &    \\
$2d$   &  11.511  & 11.499    &  & && 11.549   &  11.538   &    \\
$3p$   & 11.606   &   11.590  &   & &&  11.645  &  11.629   &  \\
\bottomrule
\end{tabular}
\caption{Masses of bottom diquark-antidiquark states in units of GeV, calculated from non-relativistic (NR) and relativistic (R) Lippmann-Schwinger equation. $\Delta$ is the absolute value of the relative percentage difference between our findings and EFG results.}
\label{Table.bqbq_bsbs}
\end{table}

We have solved both non-relativistic and relativistic form of Lippmann-Schwinger integral equation and our results indicate that the relativistic effect leads to a small reduction in the mass of heavy tetraquarks. These relativistic corrections are due to relativistic free propagator and decrease the masses by less than $2\,\%$ for charm and less than $0.2\,\%$ for bottom tetraquarks.
For both relativistic and non-relativistic calculations the same form of $D\bar{D}$ potential given by Eq. (\ref{eq.FTV}) is used.
The absolute value of the relative percentage difference between our results for $1s,1p, 2s, 1d$ and $2p$ states with those of previous studies by Ebert, Faustov, and Galkin (EFG) reported in Refs. \cite{ebert2006masses,Ebert_EPJC58,Ebert_MPLA24} 
is shown by $\Delta$ (see Tables \ref{Table.cqcq_cscs} and \ref{Table.bqbq_bsbs}).
It indicates that our results are in very good agreement with those of EFG with a relative difference estimated to be at most $2\,\%$.
Since we have ignored spin degrees of freedom in our calculations, this difference comes from the contribution of spin in the $D\bar{D}$ interaction which appears in spin-orbit, spin-spin and tensor spin-space terms \cite{Ebert_EPJC58}. 
One can say spin-dependent terms in $D\bar{D}$ interactions have a very small contribution in the masses of tetraquarks. 
While the relativistic effects leads to small reduction in the masses of tetraquarks, considering the spin degrees of freedom may leads to small reduction or increase in the masses of tetraquarks.
Carlucci et al. have predicted the masses of $1s$ state of $cq\bar{c}\bar{q}$ tetraquark with the values of $3.857$ and $3.729$ GeV and also of $bq\bar{b}\bar{q}$ tetraquark with the values of $10.260$ and $10.264$ GeV for $S\bar{S}$ and $A\bar{A}$ diquark-antidiquark contents, respectively \cite{Carlucci_EPJC57}. The masses of $1s$ state of $cq\bar{c}\bar{q}$ tetraquark is also reported by Maiani et al. as $3.723$ and $3.832$ GeV for $S\bar{S}$ and $A\bar{A}$ diquark-antidiquark contents, correspondingly \cite{Maiani_PRD71}. They have also reported the mass of $1p$ state of $cs\bar{c}\bar{s}$ tetraquark for $S\bar{S}$ diquark-antidiquark content with the value of $4330\pm70$ \cite{Maiani_PRD72}.

The calculated masses of tetraquarks should be independent of the regularization cutoff $r_c$. 
 If the largest tetraquark mass be independent of $r_c$, the lower mass states definitely would be indepenedent of the regularization cutoff.
To this aim, in Table \ref{Table.rc-dependence} we have studied the dependence of $3p$ state of bottom tetraquark $bs\bar{b}\bar{s}$ in $A\bar{A}$, as a function of the regularization cutoff. 
Clearly a regularization cutoff equal to 10 GeV$^{-1}$ is quite enough to achieve the cutoff independent results for tetraquark masses converged with at least 4 and 5 significant digits for charm and bottom tetraquarks, respectively.

\begin{table}[hbt] 
\centering
\begin{tabular}{cccccccccccc}
\toprule
$r_c$ (GeV$^{-1}$) & Tetraquark Mass (GeV)\\ \hline
\bottomrule
2.5 & 11.6316 \\
3 & 11.6274 \\
4 & 11.6263 \\
5 & 11.6322  \\
7 &  11.6285 \\
10 &  11.6288  \\
15 &  11.6288 \\
\bottomrule
\end{tabular}
\caption{The mass of bottom tetraquark $bq\bar{b}\bar{q}$ for $3p$ state (in $A\bar{A}$) as a function of the regularization
cutoff $r_c$.}
\label{Table.rc-dependence}
\end{table}

In Table \ref{Table.cqcq_experiment}, we have compared our results for the masses of charm tetraquarks with the possible experimental candidates. They are in excellent agreement with a relative difference below $0.8\,\%$.
We have also extended our calculations to higher excited states and we have successfully obtained the masses of charm and bottom tetraquarks for $3s, 2d$ and $3p$ states. To the best of our knowledge these states are calculated for the first time and we are not aware of any theoretical prediction or experimental data for these states.

\begin{table}[hbt] 
\centering
\begin{tabular}{cccccccccccc}
\toprule
& \multirow{2}{*}{state} &  \multicolumn{2}{c}{Theory}    & &  \multicolumn{2}{c}{Experiment}    \\  \cmidrule{3-4}  \cmidrule{6-7}
&     &   NR   & R  &  & Exp. candidate &   Mass                 \\ \hline 
 \multirow{9}{*}{\rotatebox[origin=c]{90}{$cq\bar{c}\bar{q}\,(S\bar{S})$}}  \\
&    &   &   & &  & 
   \multirow{3}{*}{
$
\left\{
\begin{array}{l l}
 4259 \pm 8^{+2}_{-6} & \text{\cite{PhysRevLett.95.142001}}  \\\\
 4247 \pm 12^{+17}_{-32} & \text{\cite{PhysRevLett.99.182004}}
\end{array}
\right.
$ 
}  \\ 
&  $1p$   &  4262  &  4231 & & $Y(4260)$ &\\
&   &    &   & & \\
\\
&    &   &   & &  & 
   \multirow{3}{*}{
   $
\left\{
\begin{array}{l l}
 4664\pm11\pm5 &   \text{\cite{PhysRevLett.99.142002}} \\\\
 4634^{+8+5}_{-7-8} & \text{\cite{PhysRevLett.101.172001}}
\end{array}
\right.
$
}  \\ 
& $2p$   &  4697  & 4644& & $Y(4660)$ &\\
&   &    &   & & \\ \\  
\hline
 \multirow{9}{*}{\rotatebox[origin=c]{90}{$cq\bar{c}\bar{q}\,(A\bar{A})$}}  \\
&    &   &   & &  & 
   \multirow{3}{*}{$
\left\{
\begin{array}{l l}
 4361\pm9\pm9  & \text{\cite{PhysRevLett.99.142002}} \\\\
  4324\pm24 & \text{\cite{PhysRevLett.98.212001}}
\end{array}
\right.
$}  \\ 
&  $1p$  &   4374 &  4346  &&  $Y(4360)$ &\\
&   &    &   & & \\
\\
&    &   &   & &  & 
   \multirow{3}{*}{$
\begin{array}{l l}
 4433\pm4\pm2 &  \text{\cite{PhysRevLett.100.142001} }
\end{array}
$}  \\ 
& $2s$  &   4535 & 4469 &&  $Z(4430)$ &\\
&   &    &   & & \\ 
 \bottomrule
\end{tabular}
\caption{Comparison of our numerical results for the masses of charm diquark-antidiquark states, calculated from non-relativistic (NR) and relativistic (R) Lippmann-Schwinger equation, and possible experimental candidates. The masses are in units of MeV.}
\label{Table.cqcq_experiment}
\end{table}

The theoretical uncertainties of our numerical results for tetraquark masses, in the present spin-independent formalism, arise from the effect of
the uncertainties associated with the diquark mass and also the potential parameters.
The uncertainties within the model can be evaluated and are mostly related
to the adopted approximations. The parameters of the model, such as
quark masses and parameters of the interquark potential, are rather
rigidly fixed from the analysis of meson and baryon mass spectra and
decays. The comparison of EFG predictions with data indicates
that the uncertainty of the predicted masses should be about few MeV.
For the heavy diquark masses it should be of the same order. The
uncertainty arising from the approximation of the calculated diquark
form factor $F(r)$, given in Eq. (\ref{eq.F}), is less than
$1\,\%$ \cite{Galkin-PC}. So, the overall theoretical uncertainties of our results by considering the spin effects, discussed in Sec. \ref{results}, should be about $3\,\%$.

\section{Conclusion}
Our numerical results for the masses of charm and bottom tetraquarks with regularized form of $D\bar{D}$ interactions, even by neglecting the spin degrees of freedom, are in great agreement with other theoretical predictions, especially with those reported by Ebert et al., and also by available experimental data. The effect of spin in the mass spectrum of tetraquarks can be studied by considering the realistic $D\bar{D}$ interactions in the proposed regularization method. It can be done in a three-dimensional formulation \cite{Fachruddin_PRC62}, where the total spin of $D\bar{D}$ can be treated in a helicity representation.

\section*{acknowledgments}
We would like to thank V.O.~Galkin for helpful discussions and for providing the values of strong coupling constant for different tetraquark states. This work is supported by National Science Foundation under contract NSF-PHY-1005587
with Ohio University. Partial support was also provided by the Institute of Nuclear and Particle Physics at Ohio University.

\section*{References}

\end{document}